\documentstyle{l-aa}

\begin{document}
   \thesaurus{11 (11.05.1;11.05.2;11.19.6)}

   \title{Metallicity gradients as detectors of matter distribution
	in elliptical galaxies}

   \author{A. Martinelli}

   \institute{Istituto Astronomico, Via Lancisi 29  ROMA Italy,
	e-mail martinelli@roma2.infn.it}

   \maketitle

   \begin{abstract}

In this work we discuss on the basis of
the instantaneous recycling approximation the possibility
of determining a relation between matter distribution (luminous and
dark) in elliptical galaxies and the observed metallicity gradients.

A particular case is developed in detail yielding an analitycal
expression of the above relation.

\vskip .5cm


   \end{abstract}

\section{Introduction}

\noindent
Metallicity gradients in elliptical galaxies are inferred from the variation
of metallic indices as a function of radius (see Danziger et al. 1993;
Carollo et al. 1993 and references therein for previous work on the subject).
The mechanism by which these metallicity variations could have arisen
is still unclear. The models proposed are based on the processes
developed during galaxy formation (Larson 1974 and Carlberg 1984).

Franx and Illingworth (1990) found that, for their
sample of 17 galaxies, the local (B-R) (U-R) colors are functions
of the local escape velocity $v_{esc}$ for all of the galaxies.
This result was confirmed later by Carollo et al. (1993) and 
Davies et al. (1993)
using a more direct indicator of metallicity ($Mg_{2}$) than colors
and a more
appropriate calculation of $v_{esc}$ for each galaxy.
This relation suggests that metallicity gradients
could have arisen because of the
different times of occurence of galactic
winds in different galactic regions. In fact the
galactic wind starts when the energy injected in the
interstellar medium (ISM) by supernovae (of both type I and II)
becomes equal to the binding energy of gas (Larson 1974).
Moreover, it is
reasonable to think that, when the galactic wind starts,
the SFR breaks down.
Hence, in the regions where $v_{esc}$ is low (i.e. where the
local potential is shallow), the galactic wind will develops 
earlier and the gas is less processed than
in the regions where $v_{esc}$ is higher, and the potential
well in the external regions is lower than in the internal ones.

This mechanism can produce metallicity gradients depending on the local
binding energy of gas and therefore on the matter distribution. On the basis
of this idea Martinelli et al. (1997) reproduced metallicity gradients in good
agreement with observational data, by using a Jaffe distribution for luminous
matter (Jaffe 1983) and a dark matter distributed on a diffuse halo. In their
work they do not take into account the processes occurred during galaxy
formation such as collapse or merging. They proposed a model starting with
the total mass already  present at the beginning.
This is a good assumption for elliptical galaxies since they must have
formed in a reasonably short timescale (Matteucci 1994,
Renzini 1995, Ziegler 1997).

The aim of this work is to exploit previous idea in the frame-work of
analytical models
(based on the {\it IRA} approximation) for deriving a relation between
metallicity distribution and matter distribution in elliptical galaxies
within the effective radius $R_e$.
A general formula is derived in section 2. Then in section 3 a particular
case is considered and discussed.

\section{Relation between matter and metallicity distribution}

\noindent
In the following, we consider elliptical galaxies
to have a spherical symmetry. 
We assume that at the initial time (t=0) these systems consist
only of primordial gas
and the star formation starts
at the same time in all galactic regions.

The galaxy is partitioned in zones having the shape of spheric wreath.
and thickness $\delta r$ with $\frac{\delta r}{r}<<1$. Each region is assumed
to evolve under the assumtions of istantaneous recycling,
good mixing and IMF ($\phi(m)$) constant in space and time.

\subsection{Basic equations}

\noindent
Under previous assuptions the equation for the gas evolution
in each region is (Edmundus 1990)

\begin{equation}
dg=-\alpha ds
\end{equation}
\noindent 
where $g$ is the gas mass and $\alpha ds$ is the amount of gas that does
not return into the interstellar medium when $ds$ of the interstellar
material forms into stars.

The equation for the evolution metals is:

\begin{equation}
d(gz)=(p-z)\alpha ds
\end{equation}
\noindent 
where $p$ is the {\it yield} i.e. the {\it ratio between the mass of new
ejected metals and the mass which remains locked up in low mass unevolving
stars and remnants.}

Under the assumption that $z<<1$ equation (2) has the following solution

\begin{equation}
z=-plog(\frac{g}{g(0)})
\end{equation}

\noindent
If we consider a linear star formation, i.e.

\begin{equation}
\frac{ds}{dt}=\psi(t)=\nu g
\end{equation}

\noindent
we obtain from equations (1), (2) and (4)

\begin{equation}
z=p\alpha\nu t
\end{equation}

\noindent

\begin{equation}
g=g(0)exp(-\alpha\nu t)
\end{equation}

\noindent
Equation (5) yields the metallicity in the gas. Hovewer, we are interested
to know the mean stellar metallicity.
The mass-averaged metallicity of a composite stellar population
is defined following Pagel and Patchett (1975) as:

\begin{equation}
<z>_*=\frac{1}{s(t)} \int_{0}^{s(t)}{z(s)ds}
\end{equation}

\noindent
It should be noted that, although the mass averaged
metallicity represents the real average metallicity of the composite stellar
population, what should be compared with the metallicities deduced from
observations is the luminosity-averaged metallicity.
In fact, the metallicity indices, measured from integrated spectra,
represent the metallicity of the stellar population which
predominates in the visual light.
The transformation from metallicity indices to real abundances
should then be performed through either theoretical or
empirical calibrations relating metallicity indices to real abundances.
Galactic chemical evolution models
(Arimoto and Yoshii, 1987; Matteucci and Tornamb\`e, 1987) have shown that
the two averaged metallicities are different, since metal poor
giants tend to predominate in the visual luminosity so that, in general, the
luminosity-averaged metallicity is smaller than the mass-averaged one.
However, the difference is negligible for galaxies with mass $M > 
10^{11} M_{\odot}$.

From equations (4), (5), (6) and (7) we have

\begin{equation}
\frac{<z>_*}{p}=\frac{e^{\alpha\nu t}-1-\alpha\nu t}{e^{\alpha\nu t}-1}
\end{equation}

\subsection{Relation}

\noindent
Chemical evolution in a given galactic region proceeds until galactic
wind starts.
Exactly eq. (4) holds until $t<t_{GW}$,
where $t_{GW}$ is the time of galactic wind occurence.
When $t>t_{GW}$ we must replace eq. (4) by 
$\frac{ds}{dt}=0$.  Therefore, we
can establish the following relation between metallicity in stars
and $t_{GW}$ from eq. (8)

\begin{equation}
\frac{<z>_*}{p}=\frac{e^{\alpha\nu t_{GW}}-1-\alpha\nu t_{GW}}
{e^{\alpha\nu t_{GW}}-1}
\end{equation}

\noindent
The values of $t_{GW}$ depend on $r$. In fact in the region at
galactocentric
distance $r$, $t_{GW}$ is given by the following condition:

\begin{equation}
E_{th_{SN}}(t_{GW},r)=E_{Bgas}(t_{GW},r)
\end{equation}

\noindent 
where $E_{th_{SN}}(t_{GW},r)$ is the thermal energy of the gas in that region
and $E_{Bgas}(t_{GW},r)$ is its binding energy due to the gravitational
attraction.

We want now calculate explicitly the energies appearing in eq. (10).

\subsubsection{$E_{th_{SN}}(t_{GW},r)$}

\noindent
The total thermal energy in the gas at the time $t_{GW}$ is:

\begin{equation}
E_{th_{SN}}(t_{GW},r)=\int_0^{t_{GW}}{\epsilon_{th_{SN}}(t_{GW}-x)R_{SN}(x)dx}
\end{equation}

\noindent 
where $R_{SN}(x)$ is the SN rate (either type I or II) and
$\epsilon_{th_{SN}}(t_{SN})$ is the fraction of the initial blast
wave energy which is trasferred by
the SN into the ISM as thermal energy.
A detailed description of the SN rate is available in Matteucci and
Greggio (1986).

\noindent
In the {\it IRA} approximation we can write

\begin{equation}
R_{SN}(x)=4\pi r^2\delta rF\nu g(x,r)
\end{equation}

\noindent
where

\begin{equation}
F=A\int_{M_{Bm}}^{8}{\phi (M)dM}+\int_{8}^{100}{\phi (M)dM}
\end{equation}

\noindent
and $A$ and $M_{Bm}$ are defined in Matteucci and Greggio (1986).
{\it IRA} approximation related to the SNI is questionable because
the SNI progenitors have mass  not very large and consequently
their lifetime is not negligible. Hovewer, for
small values of $t_{GW}$ the contribution of SNI is negligible with
respect to SNII (see fig. 1 of Matteucci and Greggio 1986).

Substituting (12) in (11) and by changing the variable of
integration we obtain

\begin{equation}
E_{th_{SN}}(t_{GW},r)=4\pi r^2 \delta r F\nu
\int_0^{t_{GW}}{\epsilon_{th_{SN}}(y) g(t_{GW}-y,r)dy}
\end{equation}

\subsubsection{$E_{Bgas}(t_{GW},r)$}

\noindent
We define $E_{Bgas}(t_{GW},r)$ as the energy necessary to carry the gas
in a given galactic region at galactocentric distance $r$ to $r+l$, where
$l$ is a model parameter which could be determined by means
of a dynamical approach. Hovewer, in the following dicussion,
its value is not relevant.

Let $M(r)$ be the total mass within the radius $r$. We have therefore:

\begin{equation}
E_{Bgas}(t_{GW},r)=4\pi r^2 \delta r G g(r,t_{GW})
\int_r^{r+l}{\frac{M(r')}{r'^2}dr'}
\end{equation}

\noindent
Substituting eq. (15) and (14) in eq. (10), by using eq. (6) and by
deriving with respect to $r$ we obtain finally

\begin{equation}
\frac{M(r+l)}{(r+l)^2}-\frac{M(r)}{r^2}=\frac{F\nu}{G}\frac{d}{dr}
\int_0^{t_{GW}}{\epsilon_{th_{SN}}(x)e^{\alpha \nu x}dx}
\end{equation}

\noindent
This last equation provides a relation between matter distribution
and $t_{GW}$.
On the other hand we can invert eq. (9) to have $t_{GW}$ as a function
of $<z>_*$. By substituting this last relation in (16), we obtain
the relation between matter distribution and metallicity distribution.

\section{Discussion and conclusions}

\noindent
The integral in equation (16) depends on $r$ via $t_{GW}$ and 
$\epsilon_{th_{SN}}(x)$. Hovewer if we adopt the Larson (1974)
prescription for $\epsilon_{th_{SN}}(x)$ ($\epsilon_{th_{SN}}(x)=
0.1\epsilon_0=0.1\:10^{51} erg$) the dependence is only
on $t_{GW}$ and we obtain from eq. (16)

\begin{equation}
\frac{M(r+l)}{(r+l)^2}-\frac{M(r)}{r^2}=\frac{0.1F\epsilon_0}{\alpha G}
e^{\tau}\tau'
\end{equation}

\noindent
where $\tau=\alpha \nu t_{GW}$ and the prime denotes the derivation
with respect to $r$

By deriving equation (9) with respect to $r$ we obtain

\begin{equation}
e^{\tau}\tau'=\frac{(e^{\tau}-1)^2}{\tau-1+e^{-\tau}}\frac{<z>_*'}{p}
\end{equation}

\noindent
The dependence of $<z>_*$ on $r$ can be reasonably chosen as

\begin{equation}
<z>_*=\beta r^{-\gamma}
\end{equation}

\noindent
in accordance with many observational data (Carollo et al. 1993,
Davies et al. 1993) in the range $0.1<\frac{r}{R_e}<1$.
The quantities $\beta$ and $\gamma$ are determined experimentally
and their values strongly change from galaxy to galaxy.

\noindent
Therefore equation (9) becomes

\begin{equation}
\frac{<z>_*'}{p}=-\frac{\gamma}{r} \frac{<z>_*}{p}=
-\frac{\gamma}{r} \frac{e^{\tau}-1-\tau}{e^{\tau}-1}
\end{equation}

\noindent
Substituting eq. (20) in (18) we obtain from eq. (17)

\begin{equation}
\frac{M(r+l)}{(r+l)^2}-\frac{M(r)}{r^2}=-\frac{0.1F\epsilon_0 \gamma}{\alpha G}
\frac{e^{\tau}-1}{r}\frac{e^{\tau}-1-\tau}{\tau-1+e^{-\tau}}
\end{equation}

\noindent
where $\tau$ is related to $r$ by the following equation

\begin{equation}
1-\frac{\beta}{p}r^{-\gamma}=\frac{\tau}{e^{\tau}-1}
\end{equation}

\noindent
Equations (21) and (22) can be used to obtain informations on the
matter distribution in the following way. We
consider a given value of $r$. By eq. (22) we can calculate
the corresponding value of $\tau$ (this is possible because the
function in eq. (22) is bijective).
By substituting this value in eq. (21) we obtain the difference
$\frac{M(r+l)}{(r+l)^2}-\frac{M(r)}{r^2}$.

We observe that the function
$f(\tau)=\frac{(e^{\tau}-1)(e^{\tau}-1-\tau)}{\tau-1+e^{-\tau}}$
is always positive.
Hence we can conclude from equations
(21) and (22) that $M(r)$ increases less than $r^2$ and therefore
the total mass density $\rho_{Tot} (r)=\rho_{Lum} (r)+\rho_{Dark} (r)$
decreases more rapidly than $r^{-1}$.
This result is true for all the elliptical galaxies for which
metallicity gradient is negative (i.e. $\gamma >0$).
Previous result is in agreement with a Jaffe space distribution
for the luminous matter (Jaffe 1983)

\begin{equation}
\rho_{Lum}(r)=\frac{1}{r^2(1+r)^2}
\end{equation}

\noindent
where $r$ is normalized to 1 at the radius containing half the
total emitted light in the space. This radius equals $r_e/0.763$.
Therefore, in the range $0.1R_e<r<R_e$, $\rho_{Lum}(r)$ decreases
as $r^{-2}$ (i.e. more rapidly than $r^{-1}$)

To obtain quantitative information from equations (21) and (22)
we should calculate the parameters entering in those formulas.
We must refer to the abundances of any $\alpha-$element
because they
are produced on short time scales and hence {\it IRA} approximation
is good. On the other hand the values of $\gamma$ strongly change
from galaxy to galaxy and they are also depending on the
calibration formula to compute $\alpha-$element abundance
(for example $Mg$)
from the observed indices.

In a future work we intend to use the above relation
for deriving
constraints on the presence of dark matter in elliptical
galaxies within the effective radius, since the mass $M(r)$
appearing in our formula includes both dark and luminous matter.


\begin{thebibliography}{}

\bibitem[]{} Arimoto N. and Yoshi Y.:1987, Astron. Astrophys. 173, 23
\bibitem[]{} Carlberg R.: 1984, Apj 286, 404
\bibitem[]{} Carollo C. M., Danziger I.J. and Buson L.:1993, MNRAS 265, 553
\bibitem[]{} Davies R. L., Sadler E. M. and Peletier R. F.:1993, MNRAS 262, 650
\bibitem[]{} Danziger I.J., Carollo C. M., Buson L., Matteucci F.
and Brocato E.: 1993, in {\it "Structure, Dynamics and Chemical Evolution
of Elliptical Galaxies"} eds. I.J. Danziger et al., ESO/EIPC Workshop, p.399.
\bibitem[]{} Edmundus M.G.: 1990 MNRAS 246, 678
\bibitem[]{} Franx M. and Illingworth G.:1990, Apj 359, L41
\bibitem[]{} Jaffe W.:1983, MNRAS 202, 995
\bibitem[]{} Larson R. B.:1974, MNRAS 166, 585
\bibitem[]{} Martinelli A., Matteucci F. and Colafrancesco S.:1997
MNRAS submitted
\bibitem[]{} Matteucci, F. and Greggio, L.:1986, Astron. Astrophys. 154, 279
\bibitem[]{} Matteucci F. and Tornamb\'e A.:1987, Astron. Astrophys. 185, 51
\bibitem[]{} Matteucci F.:1994, Astron. Astrophys. 288, 57
\bibitem[]{} Pagel B. E. J., Patchett B. E.:1975, MNRAS 172, 13
\bibitem[]{} Renzini A. 1995 IAU Symp. 164 {\it Stellar Populations}, ed. P. C>
\bibitem[]{} Tinsley B.M.: 1980 Fund. Cosmic Phys. 5, 287
\bibitem[]{} Ziegler B. L.:1997, in press.

\end{thebibliography}
\end{document}